\documentclass[twocolumn,floatfix,showpacs,prb,superscriptaddress,10pt]{revtex4-1}
\usepackage{graphicx}
\usepackage{amsmath}
\usepackage{amssymb}
\usepackage{bm}
\usepackage{hyperref}
\usepackage{multirow}
\usepackage{color}
\usepackage{amsbsy}

\newcommand{\pd}{{\phantom\dag}}

\begin{document}

\title{Triple point fermions in a minimal symmorphic model}

\author{I. C. Fulga}
\affiliation{Department of Condensed Matter Physics, Weizmann Institute of Science, Rehovot 76100, Israel}
\affiliation{Institute for Theoretical Solid State Physics, IFW Dresden, 01171 Dresden, Germany}

\author{Ady Stern}
\affiliation{Department of Condensed Matter Physics, Weizmann Institute of Science, Rehovot 76100, Israel}

\date{\today}
\begin{abstract}

Gapless topological phases of matter may host emergent quasiparticle excitations which have no analog in quantum field theory. This is the case of so called triple point fermions (TPF), quasiparticle excitations protected by crystal symmetries, which show fermionic statistics but have an integer (pseudo)spin degree of freedom. TPFs have been predicted in certain three-dimensional non-symmorphic crystals, where they are pinned to high symmetry points of the Brillouin zone. In this work, we introduce a minimal, three-band model which hosts TPFs protected only by the combination of a $C_4$ rotation and an anti-commuting mirror symmetry. 
Unlike current non-symmorphic realizations, our model allows for TPFs which are anisotropic and can be created or annihilated pairwise. It provides a simple, numerically affordable platform for their study.

\end{abstract}
\maketitle

\section{Introduction}

Beyond their prospects for technological applications, topological phases of matter offer a new means of exploring the fundamental theories of quantum physics \onlinecite{Beenakker2016}. Topologically non-trivial systems can host emergent quasiparticle excitations which go beyond the classification of elementary particles: the standard model of particle physics \onlinecite{Gaillard1999}. For instance, while standard model electrons are massive Dirac fermions, the electronic structure of topological semimetals such as TaAs \onlinecite{Huang2015, Yang2015}, NbAs \onlinecite{Xu2015}, and TaIrTe$_4$ \onlinecite{Koepernik2016, Belopolski2016} has been shown to host massless Weyl fermions. They are formed as protected crossings of two energy bands, momentum-space topological defects characterized by an effective Hamiltonian of the form ${\bf k}\cdot{\boldsymbol \sigma}$ \onlinecite{Nielsen1983, Murakami2007, Wan2011, Yang2011, Yan2016}. Here, ${\bf k}$ is the 3D momentum vector and ${\boldsymbol \sigma}$ is the vector of three Pauli matrices, which parametrize an effective pseudospin-$1/2$ degree of freedom.

In a recent work, Bradlyn et al. \onlinecite{Bradlyn2016} have found that certain solid state systems may have fermionic quasiparticles beyond those classified in quantum field theory: Dirac, Weyl, and Majorana fermions, all of which have a half-integer spin. Specifically, they showed that non-symmorphic crystal symmetries can lead to topologically protected triple band crossings, effectively realizing fermionic excitations that have an \emph{integer} pseudospin degree of freedom. Such quasiparticles, which we refer to as triple point fermions (TPF), have an associated low energy Hamiltonian of the form
\begin{equation}\label{eq:htpf}
 H_{\rm TPF} = {\bf k} \cdot {\bf S},
\end{equation}
where ${\bf S}$ is the vector of the three spin-1 matrices, $(S_j)_{kl}=-i \varepsilon_{jkl}$, and $\varepsilon_{jkl}$ is the fully anti-symmetric Levi-Civita tensor. As such, Eq.~\eqref{eq:htpf} describes the spin-1 generalization of a Weyl fermion. For small deviations away from ${\bf k}=0$, two bands of $H_{\rm TPF}$ disperse linearly along any momentum direction, like in a Weyl cone, while the third band remains flat.

Triple point fermions have been predicted in some materials \onlinecite{Bradlyn2016}, such as Ag$_3$Se$_2$Au and Pd$_3$Bi$_2$S$_2$, where they are protected by the combination of a three-fold rotation symmetry and two non-symmorphic glide symmetries. 
Due to this condition, they can only occur at specific points of the bulk Brillouin zone (BZ), called $P$ and $-P$.
While experimental confirmation is underway, from a theoretical perspective it is desirable to have minimal toy models which allow to study the properties of TPFs. These may include their responses to various perturbations, the structure of their associated surface states (so called topological Fermi arcs), as well as the effect of disorder.
However, the presence of multiple non-symmorphic symmetries may lead to tight-binding models with a large number of degrees of freedom, or which show other topological features beyond those desired. For instance, the toy models of Refs.~\onlinecite{Bradlyn2016, Tsuchiizu2016} show multiple Weyl points in addition to TPFs, and there are additional bands at the TPF energies, which may obscure their signatures.

In this paper, we introduce a minimal, ``hydrogen atom'' model for TPFs. We claim that the model is minimal for the following three reasons: (1) it has three bands, the minimal number required to produce Eq.~\eqref{eq:htpf}, (2) its band structure realizes the minimal number of two TPFs, which are pinned to the Fermi level and are the only band crossings in the system, and (3), the TPFs are protected only by a combination of two symmetries, a four-fold ($C_4$) rotation and an anti-commuting mirror. The first two properties make the model useful when simulating finite-size systems, and allow to study boundary physics and disorder effects efficiently. 
Due to its third property, our toy model has TPFs which are not pinned to specific points in the BZ, but can slide along the rotationally symmetric $C_4$ axis. This allows us to study the pairwise creation and annihilation of TPFs, which show a dispersion relation qualitatively different from Eq.~\eqref{eq:htpf} when they overlap in the BZ.

Our construction is based on the observation that both TPFs and Weyl cones are momentum-space topological defects. A single Weyl cone is a monopole of the Berry curvature, while a TPF is a double monopole, being in this sense topologically equivalent to a double Weyl point. Therefore, we begin by discussing a previously known toy model for double Weyl points, and then show how to modify the model in order to convert the double Weyl points into TPFs.

\section{Toy model for double Weyl points}

We begin by reviewing the model introduced in Ref.~\onlinecite{Shapourian2016} to study double Weyl points (DWP). The 3D momentum-space Hamiltonian has two energy bands and reads
\begin{equation}\label{eq:hdw}
\begin{split}
 {\cal H}_{\rm DW}({\bf k}) = & [2 - \cos(k_x) - \cos(k_y) - 2 \cos(k_z)] \sigma_z \\
 & + 2 \sin(k_x)\sin(k_y) \sigma_y \\
 & + 2 [\cos(k_x) - \cos(k_y)] \sigma_x,
\end{split}
\end{equation}
where $\sigma_i$ are Pauli matrices in the space of the two bands. Neglecting its $k_z$ dependence, the Hamiltonian Eq.~\eqref{eq:hdw} describes a stack of decoupled 2D Chern insulators in the $(x,y)$ plane, each of which is tuned to a topological phase transition between a gapped phase with a Chern number $C=0$ and one with $C=2$. The inter-layer coupling, $\cos(k_z)\sigma_z$, acts as a momentum-dependent chemical potential, bringing the 2D Chern insulators into either the $C=2$ or $C=0$ phase, depending on the value of $k_z$.

For a fixed $k_z\in(-\pi/2,\pi/2)$, the 2D $(k_x,k_y)$ slice of the BZ shows a Chern number $C=2$, while $C=0$ when $k_z\in(\pi/2,3\pi/2)$. At $k_z=\pm\pi/2$, a 2D topological phase transition takes place, which is signaled by a closing of the 2D bulk gap, or equivalently by a topologically protected band crossing in the 3D BZ. There are two such band touching points in the bandstructure of ${\cal H}_{\rm DW}$, which occur at the Fermi level, $E=0$, and at momenta ${\bf k}=(0,0,\pm\pi/2)$. The effective low energy theory close to a band crossing can be obtained by expanding Eq.~\eqref{eq:hdw} in momentum around ${\bf k}=(\delta k_x, \delta k_y, \pi/2+\delta k_z)$, leading to the double Weyl Hamiltonian
\begin{equation}\label{eq:hdwlowe}
\begin{split}
 H_{\rm DW} = & \left( \frac{\delta k_x^2+ \delta k_y^2}{2}+2\delta k_z \right) \sigma_z \\
  & + (\delta k_y^2 - \delta k_x^2) \sigma_x + 2 \delta k_x \delta k_y \sigma_y.
\end{split}
\end{equation}
Eq.~\eqref{eq:hdwlowe} describes a topological defect in momentum space. Integrating the Berry curvature on closed contours around the band touching points at ${\bf k}=(0,0,\pm\pi/2)$ gives Chern numbers $\mp2$, respectively. This is consistent with the requirement that the sum of all monopole charges throughout the BZ must vanish, which means that the minimal number of DWPs is two. Each DWP is a double monopole of Berry curvature, being topologically equivalent to two single Weyl cones. Note that unlike Weyl cones, which disperse linearly in all momentum directions, the bands of $H_{\rm DW}$ disperse linearly only in $k_z$ and are quadratic in the $k_x$ and $k_y$ directions.

The presence and position of DWPs in the bandstructure of Eq.~\eqref{eq:hdw} is a consequence of the system's symmetries. As shown in Ref.~\onlinecite{Fang2012}, the low energy structure of the band touching points, Eq.~\eqref{eq:hdwlowe}, is protected by a four-fold rotation symmetry of the form $C_4=\sigma_z$, such that
\begin{equation}\label{eq:c4dw}
 \sigma_z {\cal H}_{\rm DW}(k_x,k_y,k_z) \sigma_z = {\cal H}_{\rm DW}(k_y,-k_x,k_z).
\end{equation}
On the rotation symmetric $k_x=k_y=0$ line of the BZ, the two eigenstates of ${\cal H}_{\rm DW}$, $\psi_\pm$ have different rotation eigenvalues $\pm1$, so that their crossing may not be avoided without the symmetry being broken. If the constraint Eq.~\eqref{eq:c4dw} is weakly broken, band touching points will still persist, but each of the DWPs will split into two single Weyl cones.

Beyond their protection by $C_4$ symmetry, the position of the band crossings in energy and momentum can be seen as a consequence of other symmetries of Eq.~\eqref{eq:hdw}. Specifically, DWPs occur at opposite momenta in the BZ due to an inversion symmetry, ${\cal H}_{\rm DW}({\bf k})={\cal H}_{\rm DW}(-{\bf k})$. Moreover, they are pinned at zero energy by a reflection symmetry $R_{xy}=\sigma_x$ which maps $x\leftrightarrow y$ and anti-commutes with the Hamiltonian:
\begin{equation}\label{eq:rxydw}
 \sigma_x {\cal H}_{\rm DW}(k_x,k_y,k_z) \sigma_x = -{\cal H}_{\rm DW}(k_y,k_x,k_z).
\end{equation}
Due to the anti-commutation relation of Eq.~\eqref{eq:rxydw}, ${\cal H}_{\rm DW}$ shows an effective chiral symmetry \onlinecite{Altland1997} on the 2D mirror symmetric plane of the BZ, $k_x=k_y$. This means that at any point on the mirror plane, its two bands $\psi_\pm$ must have opposite energies, $E_+=-E_-$, and any band crossing must occur at the Fermi level, $E=0$. In the following, $C_4$ rotation and mirror symmetry will also be responsible for the protection of triple point fermions.

\section{From double Weyl points to triple point fermions}

\begin{figure}[tb]
 \includegraphics[width=0.95\columnwidth]{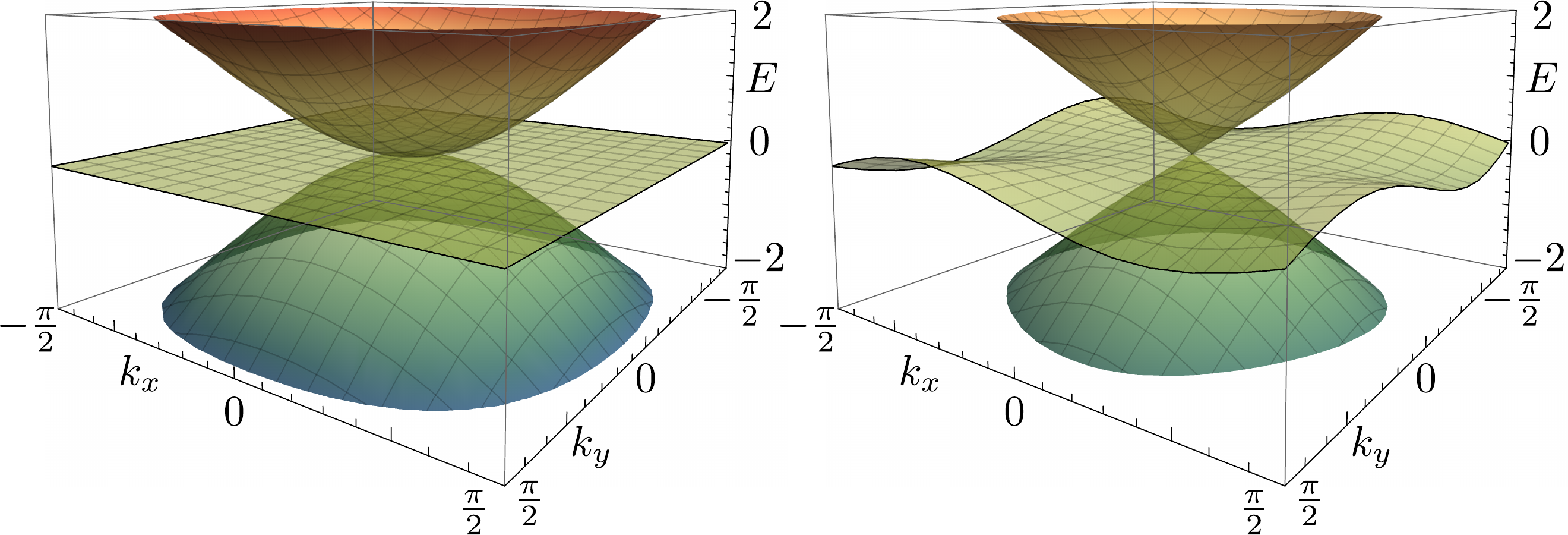}
 \caption{Closeup of the bandstructure of ${\cal H}_{\rm TPF}$ at a fixed value of $k_z=\pi/2$, using the tight-binding version of the coupling terms Eq.~\eqref{eq:lambdapm} (see text). Left panel: for $\lambda_\pm=0$ the Hamiltonian is block diagonal and the bandstructure consists of a double Weyl point intersected by a flat band. Right panel: turning on the coupling terms (here using $\rho=1$, $\phi=0$) converts the quadratically dispersing double Weyl point into a linearly dispersing triple point fermion.\label{fig:dwtotpf}}
\end{figure}

The two bands of the tight-binding model Eq.~\eqref{eq:hdw} show protected crossings, which are monopoles of the Berry curvature with charge $\pm2$, just like the TPFs of Eq.~\eqref{eq:htpf}. This fact suggests that it may be possible to deform the DWPs into TPFs by adding a third band to ${\cal H}_{\rm DW}$ in a way which preserves the total charge of the band crossing. To this end, we begin by forming a $3\times 3$ tight-binding model, in which the first $2\times 2$ block is identical to Eq.~\eqref{eq:hdw}, while the third band is chosen to be flat and positioned at zero energy. The momentum-space Hamiltonian reads
\begin{equation}\label{eq:htp}
 {\cal H}_{\rm TPF}= \begin{pmatrix}
              \multicolumn{2}{c}{\multirow{2}{*}{${\cal H}_{\rm DW}$}} & \lambda^\dag_+ \\
              \multicolumn{2}{c}{}&\lambda^\dag_-\\
              \lambda_+^\pd & \lambda_-^\pd & 0
             \end{pmatrix},
\end{equation}
where $\lambda_\pm$ are terms coupling the flat band eigenstate, $\psi_0$, to those of the DWPs, $\psi_\pm$. In general, $\lambda_\pm$ may be functions of momentum.
When setting $\lambda_\pm=0$, the Hamiltonian Eq.~\eqref{eq:htp} is block diagonal and the flat band crosses the DWPs at $E=0$, but the dispersion of the $\psi_\pm$ states remains quadratic along the $k_x$ and $k_y$ directions (see Fig.~\ref{fig:dwtotpf}, left panel).

Our aim is to increase the strength of the coupling terms $\lambda_\pm$ from 0 in such a way as to turn the quadratically dispersing DWPs into linearly dispersing pseudospin-1 fermions. This requires ensuring, on the one hand, that no anti-crossing is possible when $\lambda_\pm\neq0$, and on the other hand that all the three bands still touch at the same point in the BZ. As we will show, both of these requirements can be met by adapting the rotation and mirror symmetries of Eqs~\eqref{eq:c4dw} and \eqref{eq:rxydw} to the three-band setting of ${\cal H}_{\rm TPF}$.

First, we use a modified four-fold rotation symmetry acting in the 3-band space, $\widetilde{C}_4$, to forbid any anti-crossing. Since the original Hamiltonian  ${\cal H}_{\rm DW}$ had two bands with rotation eigenvalues $\pm1$ on the $k_x=k_y=0$ line of the BZ, we make the choice
\begin{equation}\label{eq:c4matrix}
 \widetilde{C}_4= \begin{pmatrix}
            1 &  0 & 0 \\
            0 & -1 & 0 \\
            0 &  0 & i
           \end{pmatrix},
\end{equation}
with
\begin{equation}\label{eq:c4tpf}
 \widetilde{C}^\pd_4 {\cal H}_{\rm TPF}(k_x,k_y,k_z) \widetilde{C}^\dag_4 = {\cal H}_{\rm TPF}(k_y,-k_x,k_z),
\end{equation}
similar to Eq.~\eqref{eq:c4dw}. Now for $k_x=k_y=0$, the $\psi_\pm$ states will have rotation eigenvalues $\pm1$, while the flat band state $\psi_0$ has an eigenvalue $i$. Therefore, any coupling terms $\lambda_\pm$ which obey the constraint Eq.~\eqref{eq:c4tpf} will not lead to an anti-crossing.

So far, respecting the $\widetilde{C}_4$ symmetry is insufficient to produce TPFs, even if the three bands are orthogonal on the rotation-invariant line. For instance, changing the energy of the flat band away from $E=0$ would destroy the triple band crossing while still respecting Eq.~\eqref{eq:c4tpf}. To pin the band touching point to the Fermi level, we use a modified reflection symmetry similar to Eq.~\eqref{eq:rxydw}, $\widetilde{R}_{xy}$, of the form
\begin{equation}\label{eq:rxymatrix}
 \widetilde{R}_{xy} = \begin{pmatrix}
            0 &  1 & 0 \\
            1 &  0 & 0 \\
            0 &  0 & 1
           \end{pmatrix},
\end{equation}
with
\begin{equation}\label{eq:rxytpf}
 \widetilde{R}_{xy} {\cal H}_{\rm TPF}(k_x,k_y,k_z) \widetilde{R}_{xy} = -{\cal H}_{\rm TPF}(k_y,k_x,k_z).
\end{equation}
As in the case of ${\cal H}_{\rm DW}$, the anti-commuting nature of the mirror symmetry constraint Eq.~\eqref{eq:rxytpf} means that on the $k_x=k_y$ reflection invariant plane of the BZ, the Hamiltonian ${\cal H}_{\rm TPF}$ has chiral symmetry. This constrains the bandstructure of Eq.~\eqref{eq:htp} in two important ways. First, since ${\cal H}_{\rm TPF}$ has only 3 bands, then chiral symmetry pins the flat band to $E=0$ throughout the mirror invariant plane, while the other two must exist at opposite energies $\pm E$. Second, in the presence of chiral symmetry any band crossing occurring at $k_x=k_y$ must happen at $E=0$ and must involve all three bands.

Taken together, the symmetry relations of Eqs.~\eqref{eq:c4tpf} and \eqref{eq:rxytpf} lead to protected TPFs in the bandstructure of the three-band Hamiltonian Eq.~\eqref{eq:htp}. To see this, we consider generic, momentum-dependent coupling terms $\lambda_\pm({\bf k})$ and impose the two symmetries. 
The $\widetilde{C}_4$ rotation symmetry leads to coupling terms which obey
\begin{equation}\label{eq:lambdac4}
 \lambda_\pm(k_x,k_y,k_z)=\pm i\lambda_\pm(k_y,-k_x,k_z),
\end{equation}
while the anti-commuting mirror requires
\begin{equation}\label{eq:lambdarxy}
 \lambda_\pm(k_x,k_y,k_z)=- \lambda_\mp(k_y,k_x,k_z).
\end{equation}

Given Eqs.~\eqref{eq:lambdac4} and \eqref{eq:lambdarxy}, the coupling terms become $k_z$-independent to leading order in momentum, taking the form
\begin{equation}\label{eq:lambdapm}
 \lambda_\pm = \rho \,e^{i\left( \phi\pm \frac{\pi}{4} \right)}\,(k_x\pm ik_y), \\
\end{equation}
where $\rho$ and $\phi$ are real-valued constants. On the level of the tight-binding Hamiltonian ${\cal H}_{\rm TPF}$, we use off-diagonal terms of the form Eq.~\eqref{eq:lambdapm}, replacing $k_{x}\rightarrow\sin(k_x)$ and $k_{y}\rightarrow\sin(k_y)$.

Close to the band crossing points, the linear momentum dependence of the coupling terms [Eq.~\eqref{eq:lambdapm}] becomes stronger than the original, quadratic dispersion of the two DWPs, such that they are converted into two TPFs (see Fig.~\ref{fig:dwtotpf}, right panel). Expanding ${\cal H}_{\rm TPF}$ to lowest order around ${\bf k}=(\delta k_x, \delta k_y, \pi/2+\delta k_z)$ yields
\begin{equation}\label{eq:htpkdots}
 U{\cal H}_{\rm TPF}({\boldsymbol\delta}{\bf k})U^\dag \simeq 2
 \begin{pmatrix}
         0 & -i\delta k_z & -A^*\delta k_y \\
        i\delta k_z & 0 & A^*\delta k_x \\
         -A\delta k_y & A\delta k_x & 0 \\
       \end{pmatrix},
\end{equation}
with $A=\rho e^{i\phi}/\sqrt{2}$ a complex number, and with the basis change implemented by the unitary matrix
\begin{equation}\label{eq:basischange}
 U=\frac{1}{2}\begin{pmatrix}
            1-i &  1+i & 0 \\
            1+i & 1-i & 0 \\
            0 &  0 & 2
           \end{pmatrix}.
\end{equation}

Setting $\rho=\sqrt{2}$ and $\phi=\pi/2$, such that $A=i$ in Eq.~\eqref{eq:htpkdots}, reproduces the isotropic TPF Hamiltonian mentioned in the introduction, Eq.~\eqref{eq:htpf}, which is a double monopole of Berry curvature. Note however that, unlike in non-symmorphic realizations, the band crossing Eq.~\eqref{eq:htpkdots} can be made anisotropic while maintaining its topological properties, by changing the magnitude of $A$. This can be shown by computing the dispersion of the three bands away from ${\boldsymbol\delta}{\bf k}=0$, leading to $E_0=0$ for the flat band and
\begin{equation}\label{eq:etpfpm}
 E_\pm = \pm 2\sqrt{|A|^2 \left(k_x^2+k_y^2\right)+ k_z^2}
\end{equation}
for the two dispersing bands. According to Eq.~\eqref{eq:etpfpm}, the three bands are always non-degenerate for ${\boldsymbol\delta}{\bf k}\neq0$, no matter the value of $A\neq0$. As such, there can be no other band crossings in the vicinity of the TPF, and the latter will remain a double monopole of Berry curvature, but may show different velocities in the $k_z$ and $k_{x,y}$ directions, depending on the magnitude of $A$.

\begin{figure}[tb]
 \includegraphics[width=0.65\columnwidth]{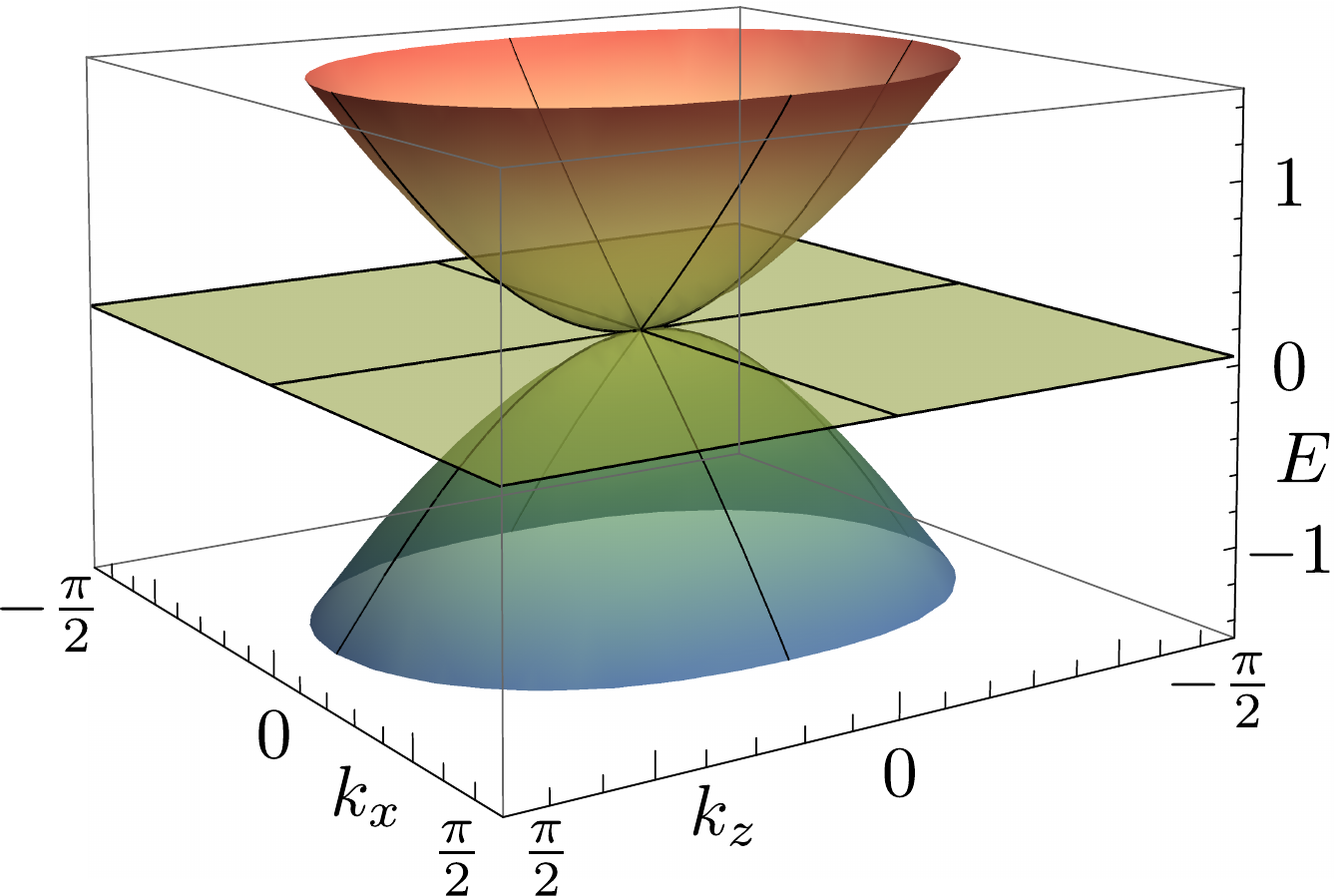}
 \caption{Closeup of the bandstructure of ${\cal H}_{\rm TPF}$ at fixed $k_y=0$, using $\rho=1$, $\phi=0$, and including the perturbation Eq.~\eqref{eq:shifttpf} with $s=2$. The two TPFs overlap at the $\Gamma$ point of the BZ, leading to a dispersion which is quadratic along $k_z$, but linear in $k_x$ and $k_y$.\label{fig:overlap}}
\end{figure}

Beyond the possibility for anisotropic TPFs, the toy model ${\cal H}_{\rm TPF}$ also allows the band crossings to move in the BZ. In current non-symmorphic models, the protecting symmetries pin the TPFs to the $P$ and $-P$ points of the 3D BZ. In the Hamiltonian Eq.~\eqref{eq:htp} however, they are only constrained to lie on the $k_x=k_y=0$ rotationally invariant line. Therefore, adding a term of the form
\begin{equation}\label{eq:shifttpf}
 \Delta{\cal H} = {\rm diag}(s,-s,0),
\end{equation}
which respects both the $\widetilde{C}_4$ and $\widetilde{R}_{xy}$ symmetries, will shift the TPFs in the $k_z$ direction. For $s>0$ the band crossings will move towards $k_z=0$ and eventually overlap at $s=2$, at which point the dispersing bands become quadratic in $k_z$ while remaining linear in $k_x$ and $k_y$ (see Fig.~\ref{fig:overlap}). For values of $s>2$ the TPFs are annihilated, and the topological semimetal is converted into a trivial band insulator.

\section{Conclusion}

Gapless topological phases of matter host protected band crossings, such as Weyl cones and nodal lines \onlinecite{Heikkilae2015, Hyart2016, Weng2016, Zhu2016, Winkler2016}. A recently discovered class of topological semimetals shows point-like crossings of three energy bands, effectively realizing fermionic excitations with an integer pseudo-spin, in contrast to the fermions of the standard model. In this work we have introduced a toy model capturing the main features of such phases, topologically protected crossings of three energy bands, or TPFs.
Rather than relying on multiple non-symmorphic symmetries, the setup we consider shows band degeneracies that are protected only by the combination of a four-fold rotation symmetry and a mirror operation which anti-commutes with the Hamiltonian [Eqs.~\eqref{eq:c4tpf} and \eqref{eq:rxytpf}]. The resulting system has only three bands and realizes the smallest possible number of two band crossings, being in this sense minimal.

The small number of orbitals in the tight-binding model ${\cal H}_{\rm TPF}$ makes it appealing for future studies on the properties and stability of TPFs. For instance, the topological surface states associated with the bulk band crossings, so called Fermi arcs, can be readily examined in a real-space version of ${\cal H}_{\rm TPF}$ (see Appendix \ref{app:fermiarcs}). The stability of the TPFs to the presence of disorder may be captured by performing a finite-size scaling of the system's density of states, similar to existing works on Weyl points \onlinecite{Shapourian2016, Sbierski2014, Pixley2016, Sbierski2016, Trescher2016, Lepori2016, Bera2016, Roy2016}. Apart from TPFs, there are no other bands at the Fermi energy, meaning there would be no spurious contributions to the density of states, and the $3\times3$ model Eq.~\eqref{eq:htp} provides an efficient, numerically affordable platform for their study.

Moreover, by using rotation and mirror symmetries, we have obtained TPFs with features that go beyond those possible in current non-symmorphic realizations. Among them, we have highlighted the possibility of moving the band crossings in the 3D BZ, such that they can be created or annihilated pairwise. Additionally, we have found that TPFs can be anisotropic, their dispersing bands showing different velocities along different momentum directions. The latter feature means that response functions to different perturbations may be direction dependent, providing an interesting avenue for future research.

Finally, we comment on possible physical realizations of the toy model ${\cal H}_{\rm TPF}$. The topological semimetal phase is protected by a mirror symmetry which anti-commutes with the Hamiltonian, whereas in most solid state systems reflections commute. It is however possible to engineer models with constraints similar to Eq.~\eqref{eq:rxytpf} by using ultra-cold atoms trapped in optical lattices. Ref.~\onlinecite{Lepori2016a} for instance proposes to realize a double-Weyl semimetal using ultra-cold atoms trapped in a cubic lattice and subject to an artificial gauge potential. When each plaquette of the lattice is pierced by a flux $\pi$, the resulting Hamiltonian shows anti-commuting mirrors. Indeed, $\pi$-fluxes provide a natural way of engineering anti-commuting mirrors, as exemplified by the well known Hofstadter model \onlinecite{Hofstadter76} on a two dimensional square lattice. Choosing a vector potential ${\bf A} = \pi (0,x,0)$ in the Landau gauge leads to a two-site magnetic unit cell, in which horizontal hoppings have a constant value $t$, whereas vertical hoppings alternate in sign along the $x$ direction. This leads to a Hamiltonian
\begin{equation}\label{eq:hofstadter}
H = t \big[ 2\cos(k_y) \sigma_z + [1 + \cos(k_x)] \sigma_x + \sin(k_x) \sigma_y \big]
\end{equation}
at half-filling, which obeys the anti-commuting mirror $H(k_x,k_y) = - \sigma_y H(k_x, -k_y) \sigma_y$. 
It is interesting to consider whether a model such as ${\cal H}_{\rm TPF}$ could be generated in a way similar to the approach of Ref.~\onlinecite{Lepori2016a} , by using three atomic levels that are close in energy.

\begin{acknowledgements}
We would like to thank Bingai Yan for useful discussions. This work was supported by the DFG (CRC/Transregio 183, EI 519/7-1), the European Research Council under the European Union's Seventh Framework Programme (FP7/2007-2013) / ERC Project MUNATOP and the US-Israel Binational Science Foundation.
\end{acknowledgements}

\bibliography{tpf}

\clearpage
\appendix

\newpage
\section{Fermi arcs}
\label{app:fermiarcs}

\begin{figure}[tb]
 \includegraphics[width=1.0\columnwidth]{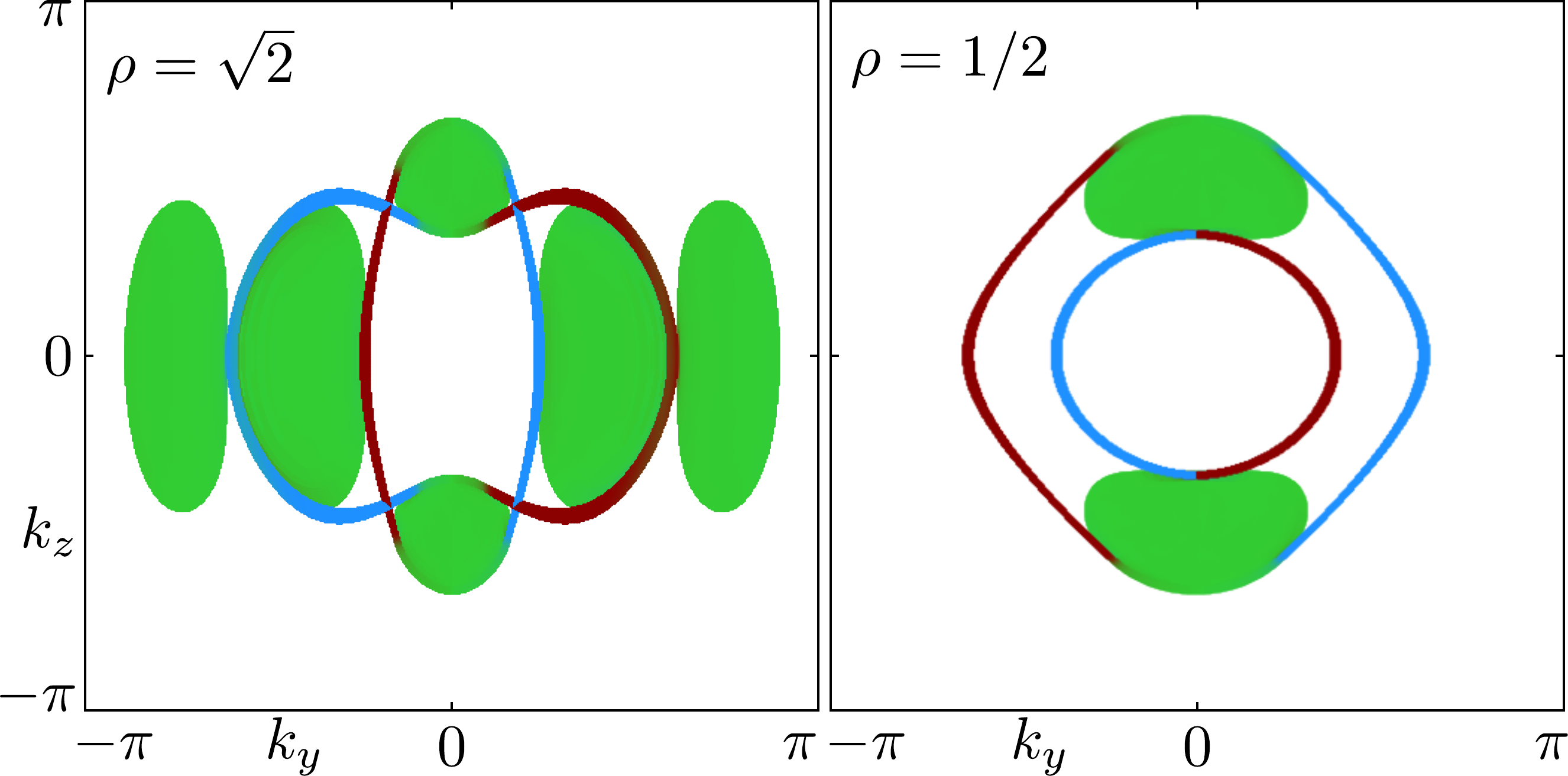}
 \caption{States with energies $0.9 \leq E \leq 1$ in the surface BZ of the model Eq.~\eqref{eq:htp}, using hard-wall boundary conditions at $x=0$ and $80$. The states localized on the top and bottom surfaces (Fermi arcs) are plotted in blue and red, respectively, while bulk states are shown in green. On each of the two surfaces, a pair of Fermi arcs emanates from the Fermi surfaces associated with the TPFs at $k_y=0$ and $k_z=\pm\pi/2$. \label{fig:fermiarcs}}
\end{figure}

In this Appendix we examine the topological surface states associated with the triple point fermions of the model Eq.~\eqref{eq:htp}. For this purpose we impose hard-wall boundary conditions in the $x$ direction, at $x=0$ and $80$ in units of the lattice constant. Close to $E=0$, the Fermi arcs overlap with the central band of $\cal{H}_{\rm TPF}$, which occupies most of the surface BZ. To distinguish between the surface and bulk contributions, we plot states with energies in the range $0.9 \leq E \leq 1$ in the surface BZ in Fig.~\ref{fig:fermiarcs}.

On each of the two surfaces (shown in blue and red), a pair of Fermi arcs connects the Fermi surfaces associated with the two TPFs at $k_y=0$ and $k_z=\pm\pi/2$, since each one is a double monopole of Berry curvature. Setting $\rho=\sqrt{2}$ in the coupling terms $\lambda_\pm$ [Eq.~\eqref{eq:lambdapm}] leads to isotropic TPFs at low energies and approximately circular TPF Fermi surfaces for $0.9\leq E\leq 1$ (see Fig.~\ref{fig:fermiarcs}, left panel). The surface BZ also shows additional Fermi surfaces due to the central band of $\cal{H}_{\rm TPF}$, which is flat close to each crossing point but acquires a nonzero dispersion away from it. Reducing the strength of the coupling terms to $\rho=0.5$ (Fig.~\ref{fig:fermiarcs}, right panel) decreases the bandwidth of the central band, such that only TPF Fermi surfaces remain. The latter are now highly anisotropic, reflecting the different velocities of the TPFs in the $k_y$ and $k_z$ directions.

\end{document}